\input{aipcheck}
\documentclass[final]{aipproc}
\layoutstyle{6x9}

\begin{document}

\title{Strange Stars : An interesting member of the compact object family}

\classification{21.65.Qr, 97.60.Jd, 97.60.Gb}
\keywords      {quark matter, equation of state, surface tension, pulsar}

\author{Manjari Bagchi}{
  address={Tata Institute of Fundamental Research, Homi Bhaba Road, Colaba, Mumbai 400005, India}
}

\author{Subharthi Ray}{
  address={Inter University Centre for Astronomy \& Astrophysics, Ganeshkhind, Pune 411007, India}
}

\author{Jishnu Dey}{
  address={Dept. of Physics, Presidency College, Kolkata 700073, India}
}

\author{Mira Dey}{
  address={Dept. of Physics, Presidency College, Kolkata 700073, India}
}

\begin{abstract}

We have studied strange star properties both at zero temperature and at finite temperatures and searched signatures of strange stars in gamma-ray, x-ray and radio astronomy. We have a set of Equations of State (EoS) for strange quark matter (SQM) and solving the TOV equations, we get the structure of strange stars. The maximum mass for a strange star decreases with the increase of temperature, because at high temperatures, the EoS become softer. One important aspect of strange star is that, surface tension depends on the size and structure of the star and is significantly larger than the conventional values. Moment of inertia is another important parameter for compact stars as by comparing theoretical values with observed estimate, it is possible to constrain the dense matter Equation of State. We hope that this approach will help us to decide whether the members of the double pulsar system PSR J0737-3039 are neutron stars or strange stars.

\end{abstract}

\maketitle

%%%%%%%%%%%%%%%%%%%%%%%%%%%%%%%%%%%%%%%%%%%%
%% MAINMATTER
%%%%%%%%%%%%%%%%%%%%%%%%%%%%%%%%%%%%%%%%%%%%
\section{Introduction}

The concept of Strange Star (SS) is not a new one. Itoh (1970) first proposed a model for it even when the QCD theory did not reach its current status. During this period Bodmer (1971) discussed about an astrophysical object collapsed from its usual state formed by nucleonic matter.  He argued that Collapsed Nuclei have radii much smaller than ordinary nuclei. Then Witten (1984) suggested that the early universe may have undergone a first order phase transition from high
temperature dense quark phase to low temperature dilute baryonic
phase, both the phases being stable with local minima. The low
temperature phase grow and gradually occupy more than half of the
volume. At this point the high temperature regions detach from
each other into isolated bubbles. Then further expansion of the
universe results in cooling in low temperature phase. These
bubbles may survive till today in the form of strange stars. He
also conjectured that the true ground state of matter is ``
Strange Quark Matter (SQM)", not $Fe^{56}$. SQM is a bulk quark
matter phase consisting of roughly equal numbers of up, down and
strange quarks plus a small number of electrons to guarantee
charge neutrality. SQM  would then have a lower energy per baryon than
ordinary nuclei and manifests in the form of strange stars.
There are several EoSs for SQM, such as the Bag model (Alcock
$et~al.$ 1986, Haensel $et~al.$ 1986, Kettner $et~al.$ 1995), the
perturbative quantum chromodynamics (QCD) model (Fraga $et~al.$
2001), the chiral chromodielectric model (Malheiro $et~al.$ 2003)
etc. However, in all these models, the effect of finite
temperature was not sufficiently studied. Such study has been
performed in the present work using the relativistic mean field
model (Dey $et~al$ 1998, Bagchi $et~al.$ 2006). In the next section, we discuss about this model.

\section{Model}

In this model, we perform a relativistic Hartree-Fock calculation for the SQM EoS, using a phenomenological inter-quark interaction,
namely a modified version of Richardson potential (Richardson, 1979). The original potential takes care of two features of the inter-quark force,
namely asymptotic freedom (AF) and confinement with the same
scale, which is not true from theoretical considerations. So we have modified the form of Richardson potential by using different scales for AF
and confinement. The scale values were obtained from baryon
magnetic moment calculations (Bagchi $et~al.$ 2004). In this model,
chiral symmetry restoration at high density is incorporated by
introducing density dependent quark masses. A temperature dependence of
gluon mass is considered in addition to its usual density dependence.
The gluon mass represents the medium effect, resulting in the screening
of the inter-quark interaction. The temperature effect through the Fermi
function is also incorporated and at non-zero temperature, free energy $F=E-TS$ is used instead of energy $E$ while calculating the EoS. It is also ensured that in SQM, the chemical potentials of the quarks satisfy ${\beta}$ equilibrium and charge neutrality conditions. The parameters of the model are adjusted in such a way that the minimum value of $E/A$ for u,d,s quark matter is less than that of Fe$^{56}$, so that u,d,s quark matter
can constitute stable stars. The minimum value of $E/A$ is
obtained at the star surface where the pressure is zero. However,
the minimum value of $E/A$ for u,d quark matter is greater than
that of Fe$^{56}$ so that Fe$^{56}$ remains the most stable
element in the non-strange world. With the obtained EsoS,
Tolman--Oppenheimer--Volkov equations for hydrostatic equilibrium
are solved to get the structures of the stars at different
temperatures (Bagchi $et~al.$ 2006). The increase of temperature softens the SQM EoS  resulting in a different curve in the mass-radius plane with a lower value of maximum mass.

\section{Surface Tension}

In Newtonian limit, surface tension is the property of the
interface between two media. But we have found that for strange
stars, where general relativity plays a significant role, surface
tension depends on the size and structure of the concerned object
(Bagchi $et~al.$ 2005) and varies from about 10 to 140 $MeV~fm^{-2}$.

Moreover, our estimated values of surface tension are
significantly larger than the conventional values (Bombaci $et~al.$ 2004) and match with the prediction that strange stars formed in the early
universe may survive evaporation (Alcock \& Olinto, 1989)  which in turn may explain the baryon-antibaryon asymmetry and the dark matter
problem according to Oaknin and Zhitnitky (2005) to yield
photons from $e^+~e^-$ annihilation.

\section{Other astrophysical implications}

We are interested in ``Double Neutron Star" (DNS) systems. Among
them, the double pulsar PSR J0737$-$3039 is most interesting
which may lead to strong tests of general relativity by
observational determination of moment of inertia of the pulsars.
It is also interesting to study the genesis of PSR J0737$-$3039B
which leads to different scenarios. We are studying the
probability of PSR J0737$-$3039B to be a strange star (Bagchi $et~al.$ 2006).

%%%%%%%%%%%%%%%%%%%%%%%%%%%%%%%%%%%%%%%%%%%
%% The following lines show an example how to produce a bibliography
%% without the help of the BibTeX program. This could be used instead
%% of the above.
%%%%%%%%%%%%%%%%%%%%%%%%%%%%%%%%%%%%%%%%%%%


\begin{thebibliography}{9}
\bibitem{itoh}N. Itoh, , Prog. Theor. Phys. \textbf{44}, 291 (1970).
\bibitem{bodmer}A. R. Bodmer, Phys. Rev. D \textbf{4}, 160 (1971).
\bibitem{witprd}E. Witten, Phys. Rev. D \textbf{30}, 272 (1984)
\bibitem{afo86} C. Alcock, E. Farhi and A. Olinto, ApJ \textbf{310}, 261 (1986).
\bibitem{han86}P. Haensel, J. L. Zdunik and R. Schaeffer,  A\&A \textbf{160}, 121 (1986).
\bibitem{ket95}C. Kettner, F. Weber, M. K. Weigel and N. K. Glendenning, Phys.Rev.D \textbf{51}, 1440. (1995).
\bibitem{fps01}E.S. Fraga, R.D., Pisarski and J. Schaffner-Bielich, Phys.Rev.D \textbf{63}, 121702 (2001).
\bibitem{mft03}M. Malheiro, M. Fiolhais, and A. R. Taurines, J. Phys. G \textbf{29}, 1045 (2003).
\bibitem{d98} M. Dey, I. Bombaci, J. Dey, S. Ray and B.C. Samanta, Phys. Lett. B \textbf{438}, 123 (1998).
\bibitem{model} M. Bagchi, S. Ray, M. Dey, J. Dey, Astron. \& Astrophys. \textbf{450}, 431  (2006).
\bibitem{rich} J. L. Richardson, Phys. Lett. B \textbf{82}, 272 (1979).
\bibitem{magmom} M. Bagchi, M. Dey,  S. Daw, J. Dey, Nucl. Phys. A \textbf{740}, 109 (2004) .
\bibitem{surf} M. Bagchi, M. Sinha, M. Dey, J. Dey and S. Bhowmick, Astron. \& Astrophys \textbf{440}, L33 (2005).
\bibitem{bom} I. Bombaci, I. Parenti, I. \& Vida\~na, Astrophys. J. \textbf{614}, 314(2004).
\bibitem{al_ol} C. Alcock, A. Olinto, Phys. Rev D \textbf{39}, 1233 (1989).
\bibitem{oz} D. H. Oaknin, A. R. Zhitnitsky, Phys. Rev. Lett. \textbf{94}, 101301 (2005).
\bibitem{dns} M. Bagchi, S. Konar, G. Bhattacharya, M. Dey and J. Dey; astro-ph/0610448.

\end{thebibliography}
\end{document}